# Local thickness and composition measurements from scanning convergent-beam electron diffraction of a binary non-crystalline material obtained by a pixelated detector

K. Nakazawa[a,*], K. Mitsuishi[b], K. Shibata[a], S. Amma[c], T. Mizoguchi[a]

[*]zawa@iis.u-tokyo.ac.jp

[a] Institute of Industrial Science, The University of Tokyo, Japan
[b] National Institute for Materials Science, Japan
[c] New Product R&D Center, AGC Ltd., Japan

Abstract:
  We measured the local composition and thickness of $SiO_2$-based glass material from diffraction. By using four dimensional scanning transmission electron microscopy (4D-STEM), we obtained diffraction at each scanning point. Comparing the obtained diffraction with simulated diffraction patterns, we try to measure the local composition and thickness. Although this method requires some constraints, this method measured local composition and thickness with 1/10 or less electron dose of EELS.

1. Introduction

  Thickness and composition of a specimen is crucial information to analyze image, diffraction, and spectrum observed by transmission electron microscopy (TEM). The thickness and composition are mainly measured by spectroscopic methods, such as electron energy loss spectroscopy (EELS) or energy dispersive X-ray spectroscopy (EDS). In EELS, the relative thickness is measured by the log-ratio method, and the composition can be determined by the ratios between core-loss intensities of each atom [1]. In EDS, the thickness and composition of the sample are simultaneously determined by the ζ factor method [2]. Since these spectroscopic methods require a large electron dose, they cannot be applied to electron-sensitive materials.

  In crystalline materials, in addition to spectroscopy, imaging and diffraction are used to measure the thickness and composition. High-angle annular dark-field (HAADF) imaging enables the thickness and column composition to be determined at the single-atom level [3–6]. The convergent-beam electron diffraction (CBED) method enables the determination of the thickness at a sub-nanometer accuracy by analyzing the intensity distribution of diffraction disks [7,8]. The difference in the thickness affects the number of scattering events, and the difference in the composition affects the scattering cross section. These differences eventually appear in the CBED patterns. By matching the CBED patterns with the library of simulated diffraction patterns, the local thickness and composition can be determined. The development of the high-performance electron pixelated detector technique has enabled the acquisition of CBED patterns easily at each scanning point with high speed. This method, which acquires diffraction patterns (reciprocal 2D) at all scanning points (real 2D), is called 4D-scanning transmission electron microscopy (STEM) (reciprocal 2D × real 2D = 4D) [9], and enables a mapping of the thickness or composition of crystalline materials [10].

  As stated above, there are alternatives to spectroscopic techniques for measuring the thickness and composition of crystalline materials. However, there are few alternatives to spectroscopic techniques for measuring the thickness and composition of non-crystalline materials [11]. In particular, the thickness and composition of non-crystalline materials have not been measured by diffraction. This is because non-crystalline materials do not have long-range order and do not make regular diffraction patterns like crystals. However, the intensity of the high angle region of diffraction increases as the thickness and the atomic number of the sample increase. Since non-crystalline materials are macroscopically isotropic, the diffractions of non-crystalline materials mainly depend on the thickness and composition of the samples. Therefore, there is a possibility that information regarding the thickness and composition can be extracted from the diffraction simultaneously.

In this article, we applied the 4D-STEM method to a BaO-SiO$_2$ glass system. By comparing the diffraction obtained by the experiment with the simulated diffraction under some constraints, we obtained both the thickness and composition maps. To verify the accuracy of these mappings, these were compared with the maps obtained by EELS and HAADF. As a result, through some constraints as this method requires, we succeed in the determination of both thickness and composition mappings with one-tenth of the electron irradiation of the EELS method.

2. Material and Methods

2.1. Procedure to determine thickness and composition

The schematic procedure used to determine the thickness and composition is shown in Fig. 1. The diffraction pattern obtained by 4D-STEM was compared with those obtained by simulation. For this comparison, using the property that diffractions of glasses have no azimuthal dependency, only the azimuthally-averaged intensity (AAI) of the diffraction pattern was compared. An AAI is obtained by dividing a diffraction pattern into regions every 10 mrad from the center and averaging each region (the top row of Fig. 1). There are three reasons why we used discrete expressions. First, in this experiment, we only can acquire discrete diffraction pattern by pixelated detector. Second, discrete expressions are easier to handle in computer. Third ,by averaging in 10 mrad increments, the effect of noise is reduced. The obtained AAIs ($AAI^{exp}$) were compared with the simulated AAI ($AAI^{sim}$) (the second and third row of Fig. 1). We measured the mean squared error (*mse*) according to the following formula,

$$ratio = \frac{\sum_r AAI^{sim}(r)}{\sum_r AAI^{exp}(r)} \quad (1)$$

$$mse = \frac{1}{n_r}\sum_r \left(ratio \times AAI^{exp}(r) - AAI^{sim}(r)\right)^2 \quad (2)$$

where $n_r$ is the number of divided regions of the AAI, $AAI(r)$ is the intensity of the AAI at radius $r$ (mrad) in the diffraction pattern. The ratio is calculated to adjust the total electron dose. Low mse means a high similarity and high mse means a low similarity between the two AAIs. We adopted the thickness and composition that produced the AAI that had the lowest mse as the composition and thickness at the scanning point (the bottom row of Fig. 1). We performed the above comparison for all scanning points and obtained the thickness and composition maps.

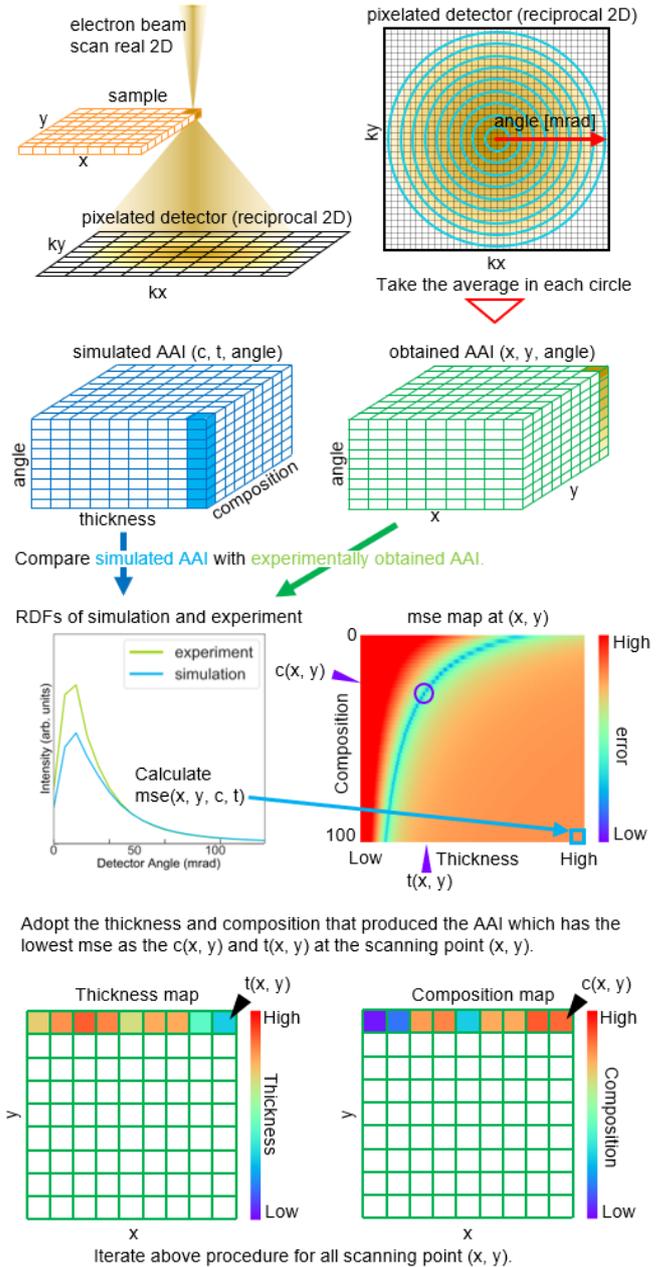

**Figure 1**. Schematic procedure to detect the thickness and composition by 4D-STEM and simulation. $x$ and $y$ are the lateral and longitudinal positions of the scanning point, respectively. $c$ and $t$ are the composition and thickness, respectively. An azimuthally-averaged intensity (AAI) is obtained by dividing a diffraction pattern into regions every 10 mrad from the center and averaging in each region. The obtained AAIs are compared with the simulated AAI and calculate mean square error (*mse*). We adopt the thickness and composition that produced the AAI that has the lowest *mse* as the composition and thickness at the scanning point.

2.2. 4D-STEM, EELS, HAADF

The silicate-based glass 27.0 BaO-73.0 $SiO_2$ (mol%) was selected for the experiment. This composition is known to separate into Si-pure phases and Ba-rich phases [12–14]. The bulk glass samples were prepared by the conventional melt-quench method. The mixed batch materials (analytical reagent grade) of $SiO_2$ and BaO were melted in an electric furnace at 1650 °C in a $Pt_{90}Rh_{10}$ crucible for 3 hours. The melt was quenched in water to produce small fragments, and then the fragments were collected, mixed and remelted in the crucible at 1680 °C for 3 hours to homogenize the glasses. The second melt was rapidly quenched by casting into two rollers with a flow of cooling water to prevent phase separation. Hereafter, this sample is referred to as the quenched sample. The obtained flakes from the quenching melt were quite transparent. Some of the flakes were annealed at 800 °C in air for 480 minutes to promote phase separation. This sample separated into a minor Si-pure phase and major Ba-rich phases at 800 °C [13,15]. Hereafter, this sample is referred to as the annealed sample. Both samples were thinned by the crushing method and the thinned samples were dispersed on a TEM grid with a carbon mesh.

We used $SiO_2$ sphere with a diameter of 1000 nm (Micromod Partikeltechnologie GmbH) to measure the precision of thickness measurement. We dispersed the spheres on a TEM grid with a carbon mesh.

STEM observation was performed using an aberration-corrected scanning transmission electron microscope (JEM-ARM200F, JEOL Ltd.) operated at an accelerating voltage of 200 kV. The convergent semi-angle was set to 20.8 mrad. We performed 4D-STEM, high-angle annular dark-field (HAADF), and EELS. We set the camera length at 25 mm to secure a large collection angle when performing 4D-STEM. The collection angle for 4D-STEM was in the range of 0–130 mrad. The 4D-STEM measurement was performed by 4D-canvas (JEOL. Ltd.). Dwell times and pixel sizes were set to 1 ms and 1.6 nm × 1.6 nm for 4D-STEM, 10 ms and 7.7 nm × 7.7 nm for HAADF and the thickness measurement by EELS, and 2 s and 20 nm × 20 nm for the measurement of the composition by EELS.

2.3. Simulation of diffraction patterns

The diffraction patterns were simulated by the multi-slice method using the Dr. Probe package [16] with form factors of Weickenmeier & Kohl[17], which incorporates the effect of inelastic scattering. The composition of the atomic structure was varied from 0 BaO-100 $SiO_2$ to 100 BaO-0 $SiO_2$ in 1 mol% increments, and the thickness was varied from 0 nm to 1000 nm in increments of 1 nm. The 0 BaO-100 $SiO_2$ glass structures were created by molecular dynamics simulation using LAMMPS code [18]. The Ba-doped glass structures were made by randomly replacing Si atoms with Ba atoms and erasing O atoms. Though this is a simple structure construction, the densities of these structures were almost equal to that from a previous experiment [15]. We simulated the AAIs of the electron diffraction patterns under the condition that the acceleration voltage was 200 kV, defocus was 0 nm, convergence semi-angle was 20.0 mrad, AAI range was from 0 to 130 mrad, and step size of AAI was 10 mrad.

3. Results and Discussion

3.1. Simulation

Figure 2 shows some of the simulated AAIs. A few electrons were scattered to a high angle. To make it easy to visualize, the intensity was multiplied by the square of the radius. Comparing samples with the same composition, a thicker sample resulted in more electrons that were scattered to a higher angle and fewer electrons scattered to a lower angle. This result was caused by the effect of multiple scattering. Comparing AAIs of the same thicknesses, a heavier sample (i.e., the more BaO in the sample) resulted in more electrons that were scattered to a higher angle and fewer electrons scattered to a lower angle. Generally, heavier atoms have a larger scattering cross section, and Ba is heavier than Si. Thus, the result arose from the large scattering cross section of Ba. Next, we compared the simulated AAIs with each other. The *mse* was calculated by Eqns (1) and (2).

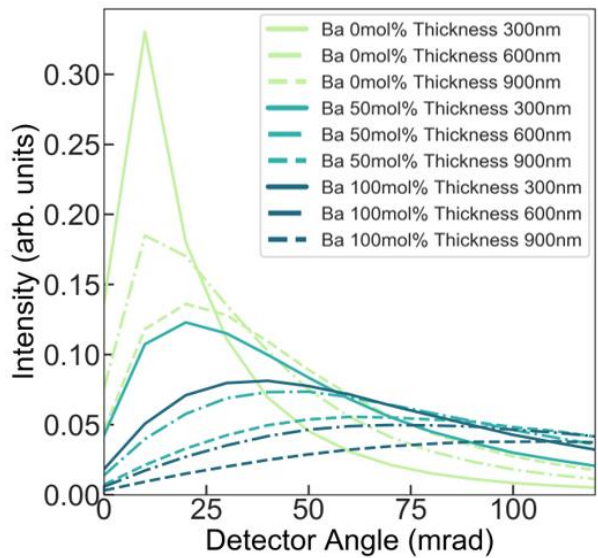

**Figure 2.** Simulated azimuthally-averaged distribution of electron scattering. The intensity is multiplied by the square of the radius to make it easy to see the high angle region. Mole fraction of Ba ranges from 0% to 100% in increments of 50%. Thickness ranges from 300 to 900 nm in increments of 300 nm. Light green, green and dark green lines correspond to the azimuthally-averaged distribution of the atomic structure with 0 mol%, 50 mol% and 100 mol% of BaO. Normal line, dashed line and dash dot lines correspond to the azimuthally-averaged distribution of the atomic structure with a thickness of 300 nm, 600 nm and 900 nm.

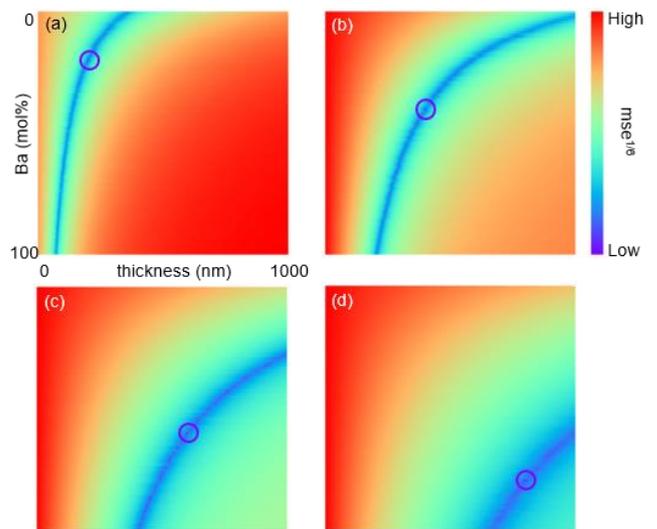

**Figure 3.** Distributions of mse when comparing the simulated AAI with each other. Red means high mse (= low similarity) and blue means low mse (= high similarity). To make it easier to visualize, the mean square error value is raised to the 1/6th power. Comparison of the AAIs with (a) 20 mol% BaO and 200-nm thickness, (b) 40 mol% BaO and 400 nm, (c) 60 mol% BaO and 600 nm, (d) 80 mol% BaO and 800 nm with the rest of the simulated AAIs. The blue circle in each image indicates the position of the compared AAI.

Figure 3 shows the *mse* distributions. These *mse* distributions were obtained by comparing an AAI with a certain composition and thickness with the rest of the simulated AAIs. To make it easier to visualize, the *mse* value was raised to the one-sixth power. For example, Fig. 3(a) was obtained by comparing the AAI of 20.0 $BaO$-80.0 $SiO_2$ and 200-nm thickness with the rest of the simulated AAIs.

In Fig. 3, the regions with a low *mse* distributed in an arc from the lower left to upper right, which showed an AAI with a certain composition and thickness was similar to those with lighter compositions and higher thicknesses, or those with heavier compositions and lower thicknesses. This result indicated that it was difficult to distinguish the effect of the scattering cross section caused by the change of composition from the effect of multiple scattering caused by the change of thickness only from the AAI. However, there was only one minimum value in the error distribution in both the composition direction and the thickness direction. Therefore, by this comparison, the combination of composition and thickness that results in similar AAIs can be obtained. There was a one-to-one correspondence between composition and thickness. Thus, if the concentration is known, the thickness can be determined uniquely from the AAI, and vice versa. In the following sections, using the relationship between thickness and composition, we determined the thickness of a sample with a known composition and determined the thickness and composition of a sample with an unknown thickness and composition under certain constraints.

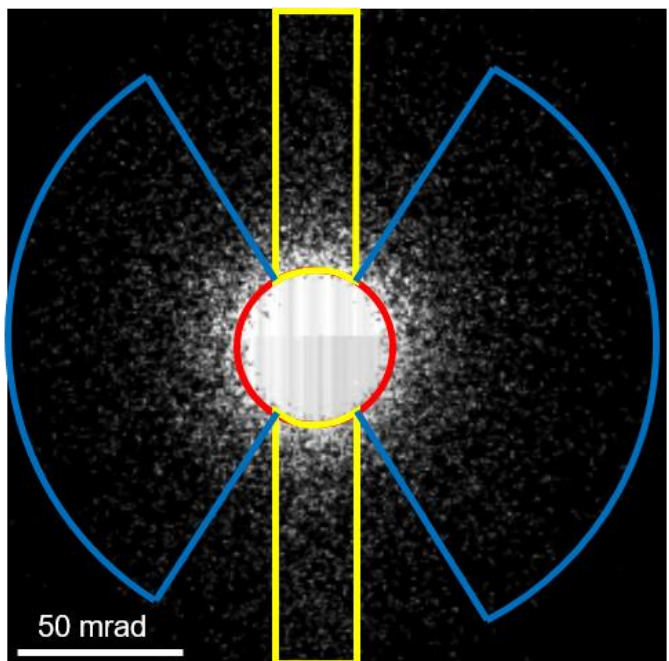

**Figure 4.** Recorded diffraction pattern of a quenched sample. The intensity at the center of the diffraction pattern (inside the red circle) is constant because of saturation. There is readout noise above and below the saturated region (depicted by yellow region). To avoid these errors, we used the blue region to calculate the AAI.

3.2. Experiment

First, we attempted to determine the thickness of the sample with a known composition by 4D-STEM and simulation. We selected a quenched glass sample, which had a uniform compositional distribution of 27.0 BaO-73.0 $SiO_2$. Figure 4 shows the experimentally observed diffraction pattern acquired by a pixelated detector. Owing to the high electron brightness, the intensity of the center of the diffraction was saturated, and readout noise was present above and below the center of the image. Thus, we used the rest of the region of the diffractions to calculate the AAI. The experimentally obtained AAIs were compared with the simulated AAIs for a composition of 27.0 BaO-73.0 $SiO_2$. The thickness and composition maps obtained by this operation are shown in Fig. 5(a) and (b). The thickness gradually increased from the top to the bottom of the sample. The same region was investigated by STEM-EELS. The relative thickness map measured by the EELS log-ratio method is shown in Fig 5(c). The thickness gradually increased from the top to the bottom of the sample, which showed good agreement with Fig.5 (a).

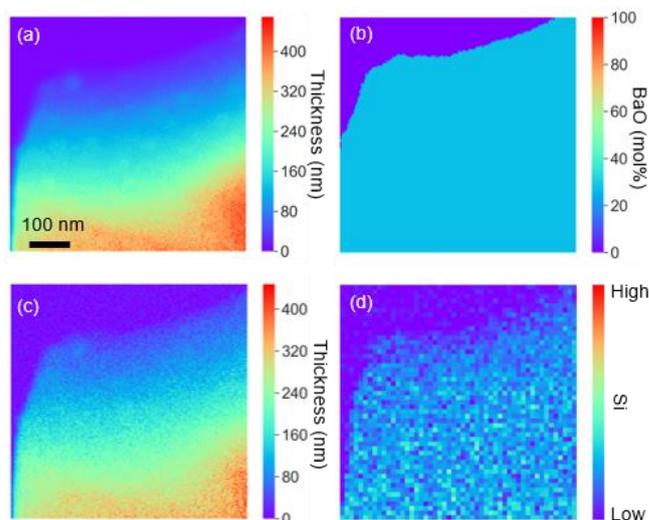

**Figure 5.** (a) Thickness map and (b) composition map detected by 4D-STEM and simulation. (c) Relative thickness map measured by EELS. Absolute thickness is calculated by multiplying by the mean free path (173 nm). (d) Composition map measured by EELS. All data are acquired from the same area in the quenched sample.

The Si map measured by the Si-$L_{2,3}$ edges (100–130.5 eV) are shown in Fig. 5(d). The Si map showed a uniform compositional distribution inside the sample. According to the nominal composition, the composition of the uniform compositional distribution was 27.0 BaO-73.0 $SiO_2$. As the gradation of the thickness map made by 4D-STEM was almost the same as that by EELS, 4D-STEM and simulation accurately determined the thickness map. Previous reports state that the mean free paths of $SiO_2$ and BaO under a large collection angle are 183 nm and 147 nm, respectively [1,19]. We adopted the weighted harmonic average of the reported mean free paths (172 nm) as the mean free path of the sample. We obtained an absolute thickness map by multiplying the relative thickness map and the mean free path of the sample. By comparing the absolute map with the thickness map obtained by the 4D-STEM method, the 4D-STEM method estimated a thickness that was 1.03-times larger than that determined by the EELS method(The detailed thickness ratio of thickness measured by 4D-STEM divided by thickness measured by EELS is shown in Fig. S1.).

In this section, we attempted to detect the thickness and composition of a sample with an unknown thickness and composition. We analyzed the annealed sample whose composition was not uniform because of phase separation.

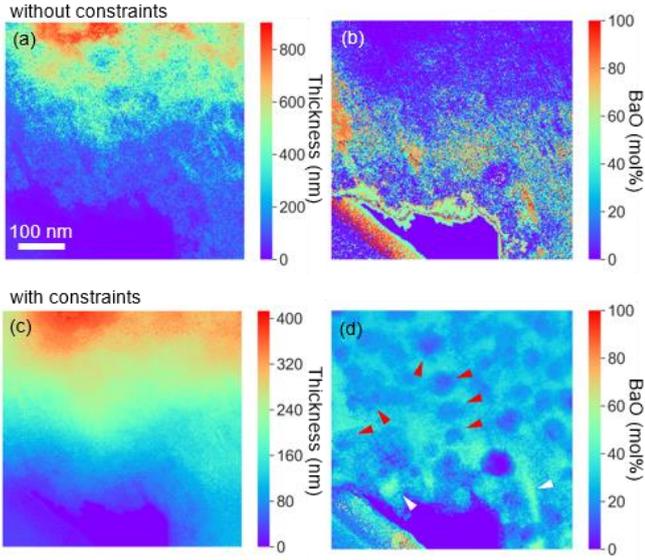

**Figure 6.** Thickness and composition maps detected by 4D-STEM and simulation (a), (b) without constraints, (c), (d) with constraints. The bottom left regions of the images are carbon meshes.

At first, we determined the composition and thickness only by comparing AAIs. The results are shown in Fig. 6(a) and (b). The signals at the bottom left region of the images were caused by the carbon mesh. The bottom middle region of the image was the vacuum, and the rest was the sample. The average composition of the results was 10.3 BaO-89.7 SiO$_2$. However, the nominal composition of this sample was 27.0 BaO-73.0 SiO$_2$. Thus, the simple comparison determined by AAIs did not yield the correct pair of thickness and composition owing to the non-uniqueness. The error in the determination might have been caused by the difficulty to distinguish the effect of the scattering cross section from the effect of multiple scattering. To prevent misidentification, it is necessary to impose constraints.

Here, to accurately extract the information of the thickness and composition from the AAIs, we imposed two constraints concerning the thickness and composition. The first was that the thickness distribution did not change steeply, namely, the thickness at one point was not largely different from the surrounding points. The other was that the average composition of the sample in the observation region was 27.0 BaO-73.0 SiO$_2$. Since there are restrictions on the combination of the composition and thickness that reproduces an AAI measured by experiment, the composition is also corrected when a constraint of the thickness is imposed. The same applies to the thickness when the constraint of the composition is imposed. The thickness and composition of each pixel were iteratively constrained by the following equations.

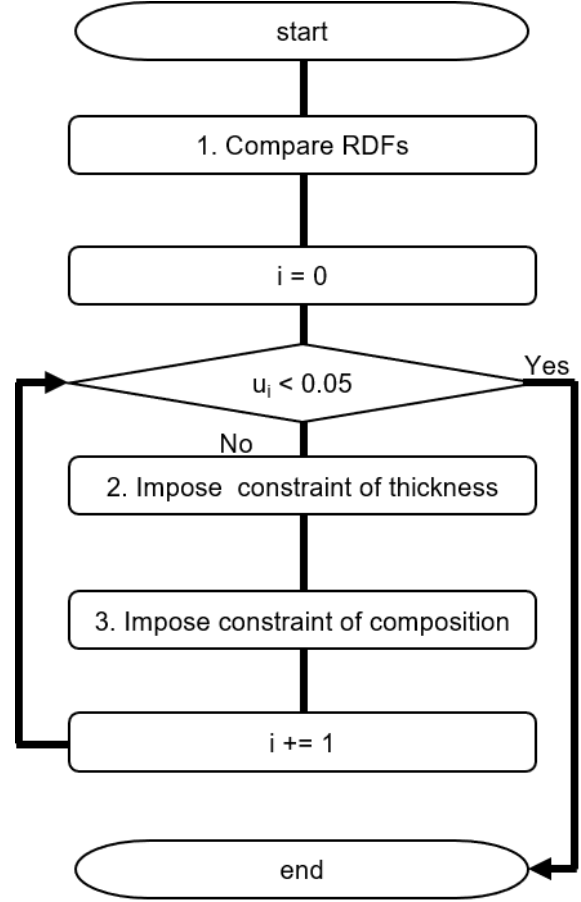

**Figure 7.** Flowchart of how to determine the thickness and composition of a sample with unknown thickness and composition by 4D-STEM and simulation.

$$t_{i+1,x,y} = t_{i,x,y} + \beta(t_{i,x,y}^{\text{ave}} - t_{i,x,y}) \quad (3)$$
$$p_{i,x,y} = t_{i,x,y} c_{i,x,y} \quad (4)$$
$$c_{i+1,x,y} = c_{i,x,y} + \beta(c^{\text{const}} - p_{i,x,y}^{\text{ave}}/t_{i,x,y}^{\text{ave}}) \quad (5)$$

where $t_{i,x,y}, c_{i,x,y}$ are the thickness and mole fraction of BaO at the position ($x$, $y$ ($x$ indicates the lateral position, $y$ indicates the longitudinal position)) at iteration number $i$. $p_{i,x,y}$ is the product of $t_{i,x,y}$ and $c_{i,x,y}$. $t_{i,x,y}^{\text{ave}}, c_{i,x,y}^{\text{ave}}, p_{i,x,y}^{\text{ave}}$ are the average thickness, composition and product in a square window with a side length of $w$ around the position ($x$, $y$). $\beta$ is the updating step parameter to prevent oscillation. Thus, $\beta$ must be smaller than 1. We used $t_{i,x,y} c_{i,x,y} / t_{i,x,y}^{\text{ave}}$ instead of $c_{i,x,y}$ to converge the volume average of the composition, rather than the area average of the composition, to $c^{\text{const}}$. In this process, we set the window size $w$ = 55 pixels (80 nm), $\beta = 0.2$ to apply 20% correction in each iteration, $c^{\text{const}} = 27.0$ to converge the average mole fraction of BaO in $w$ squares around the pixel of interest at 27 mol%(w dependence is shown in Fig S2). The determination algorithm of the thickness and composition is schematically shown in Fig. 7.

First, AAIs acquired from the experiment were compared with AAIs created by the simulation to obtain a relationship between the thickness and composition. Second, the thickness was corrected by the constraint. Third, the composition was corrected by the constraint. The second and third procedures were iterated until the updated amount of the composition fell below 0.05. The updated amount at the $i$th iteration $u_i$ was measured by the following formula,

$$u_i = \frac{\Sigma_y \Sigma_x |c_{i,x,y} - c_{i-1,x,y}|}{n_{\text{pixels}}} \quad (6)$$

where $n_{\text{pixels}}$ is the number of pixels in the image. After 14 iterations, $u_{14}$ converged to less than 0.05. (The updating rate in each iteration of composition and thickness is shown in Fig. S3, updating process of compositional mapping is shown in Fig. S4 and updating process of thickness mapping is shown in Fig. S5.)

As a result, the thickness and composition maps shown in Fig. 6(c) and (d) were acquired. The average composition was 26.4BaO-73.6SiO$_2$, which is consistent with the nominal composition. We compared these maps obtained by 4D-STEM with the thickness and composition maps measured by EELS. The Si and Ba and thickness maps measured by EELS are shown in Fig. 8(a) (b) and (e). The Si and Ba maps were created by extracting the Si-L$_{2,3}$ edge (107.6–126.1 eV) and Ba-M$_{4,5}$ edge (738.1–815.6 eV) from EELS. A HAADF image is also shown in Fig. 8(c). The dark circular contrast in the HAADF image was caused by a compositional change. As a heavier atom makes a higher intensity and there was no circular contrast in the thickness map, the dark circular region in the HAADF image contained less BaO than the other regions. Different from the HAADF image, as depicted by the red arrows, the Si and Ba EELS maps did not contain the dark circular contrast. This difference between the HAADF image and the Si or Ba EELS maps was derived from the insufficient signal-to-noise ratio of the EELS method. Here, we created another composition map by HAADF and a thickness map measured by EELS. The HAADF intensity can be roughly determined by the following equation [20,21],

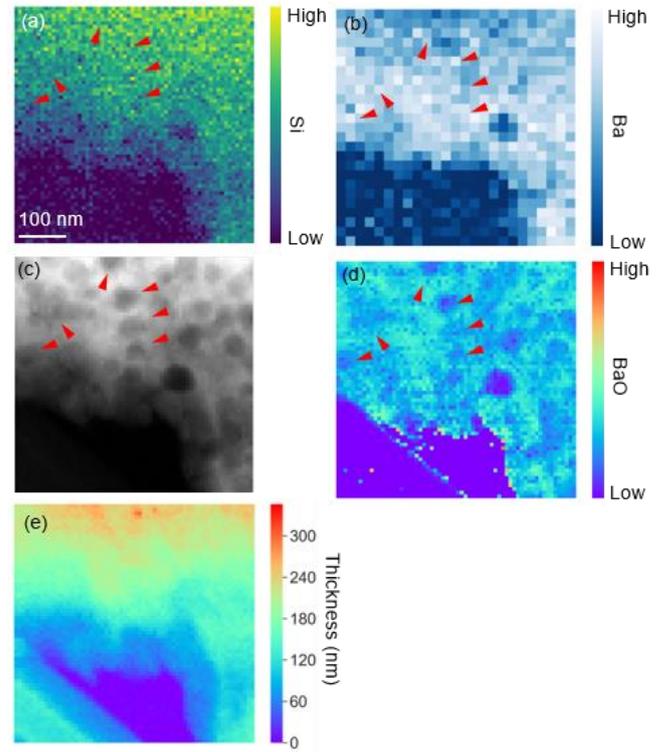

**Figure 8.** Images of the annealed sample. (a) Si map and (b) Ba map obtained by EELS. (c) HAADF image. (d) Ba map made from HAADF image and relative thickness map. Bottom left regions of the images are carbon meshes. (e) Relative thickness map measured by EELS. Absolute thickness values are calculated by multiplying by the mean free path (172 nm).

$$I^{\text{HAADF}} = I^{\text{total}} \left(1 - \exp\left(-\sum_k n_k \sigma_k t\right)\right) \quad (7)$$

where $I^{\text{HAADF}}$ is the intensity of HAADF, $I^{\text{total}}$ is the intensity of the incident electron beam, $\sigma_k$ is the Rutherford scattering cross-section of element $k$, $n_k$ is the number of elements $k$ in unit volume, and $t$ is the thickness of the sample. Rearranging the above equation, we obtain the following equation:

$$\sum_k n_k \sigma_k = -\frac{\ln\left(1 - \frac{I^{\text{HAADF}}}{I^{\text{total}}}\right)}{t} \quad (8)$$

where $\sigma_k$ is almost proportional to the square of the atomic number [22,23]. The left side of the equation is the composition weight by scattering cross-sections. Thus, the right side is proportional to the composition. The compositional map created by the above equation is shown in Fig. 8(d). This map contains all the dark circular contrast, as indicated by red arrows.

The composition map created by 4D-STEM showed good agreement with the composition map created by HAADF and EELS. The region with a low BaO concentration was consistent with each other, as depicted by the red arrows. The composition approximately ranged from 0 to 43 BaO mol%. This range was almost consistent with the phase diagram, which expected that the composition ranged from 0 to 35 BaO mol% [15]. Some of the edge regions of the sample, which are depicted by white arrows, had a high BaO mole fraction (higher than 50 mol%). This value exceeded the binodal line (around 35 mol%). This error was caused because the assumption about the composition was not valid for the edges of the sample. This method could acquire the composition map more accurately, except for the sample edges, than the method using EELS. The thickness map created by 4D-STEM also showed good agreement with the thickness map created by EELS. The thickness gradually increased from the bottom to the top of the sample. We also calculated the absolute thickness by multiplying the relative thickness map made by EELS and the mean free path. By comparing the thickness map measured by the 4D-STEM method with the absolute thickness map, the 4D-STEM method assessed a thicknesses that was 1.10-times higher than that determined by the EELS method(The detailed thickness ratio of thickness measured by 4D-STEM divided by thickness measured by EELS is shown in Fig. S1.). Although this method required two constraints, this method can qualitatively measure the thickness and composition from AAIs.

Here, to assess the precision of detecting thickness, we measured the thickness of a $SiO_2$ particle by the present 4D-STEM method. BF and HAADF images are shown in Fig. 9(a) and (b), respectively. Since, this sample is commercially available and made of pure $SiO_2$ with a sphere-shape, we imposed the constraint that the composition is 0.0 BaO-100.0 $SiO_2$. Figure 9(c) shows the thickness of the $SiO_2$ sphere measured by the 4D-STEM method. The diameter of the sphere measured along in-plane direction is 1084 nm. We calculated the thickness ratio by dividing the thickness of a sphere with a diameter of 1084 nm by the thickness measured by 4D-STEM. The result is shown in Fig. 9(d). The line profiles of Fig. 9(c) and (d) are shown Fig. S6. In

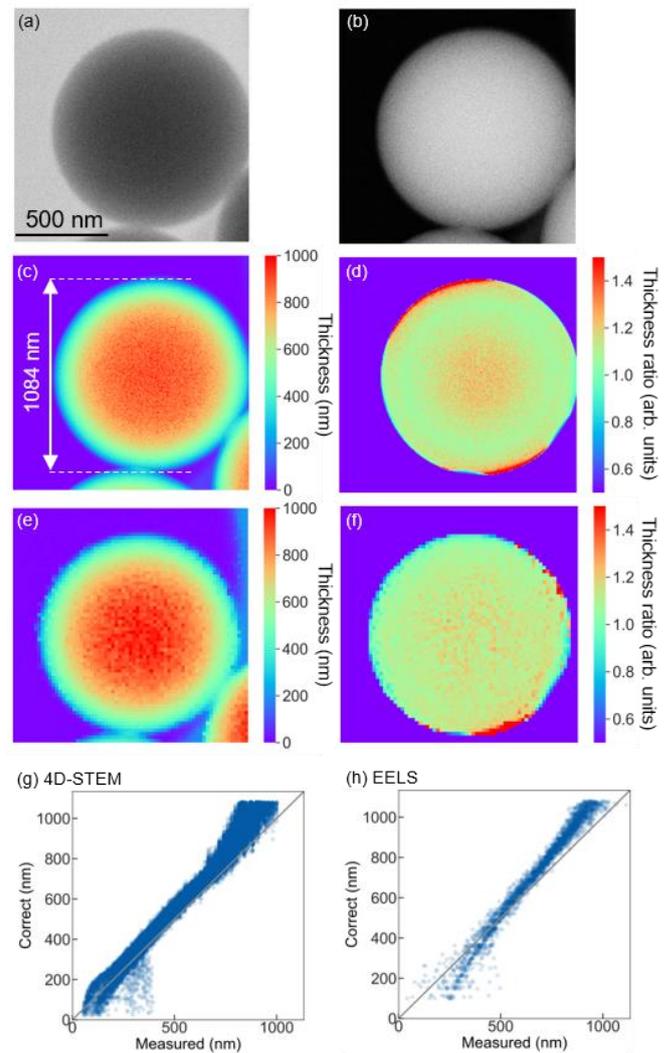

**Figure 9.** (a) BF and (b) HAADF images of SiO2 particle. (c) Thickness map measured by 4D-STEM. (d) The ratio calculated by dividing the thickness of a sphere with a diameter of 1084 nm by the thickness measured by 4D-STEM. (e) Thickness map measured by EELS. (f) The ratio calculated by dividing the thickness of a sphere with a diameter of 1084 nm by the thickness measured EELS. (g) The vertical axis shows the thickness of the assumed sphere with diameter 1084 nm, and the horizontal axis shows the thickness measured by 4D-STEM at the same position. (h) The vertical axis shows the thickness of the assumed sphere with diameter of 1084 nm, and the horizontal axis shows the thickness measured by EELS at the same position.

Fig. 9(c), thickness ratio is exceptionally high at the top left and bottom right of the sphere. This may be due to a slight distortion in shape from the sphere. The ratio is around 1.1 at the edge of the sphere (whose thickness is less than 800 nm) and 1.2 near the center of the sphere (whose thickness less than 800 nm). In the thicker area, the present method tends to underestimate the thickness. The reason of underestimation is the difference in shape between simulation and experiment. Although the sample is sphere, the simulation assumes a flat sample. In the experiment, when electrons are irradiated near the center of the sphere where is the thickest in the sphere, the electrons are scattered to surrounding thin area and the thickness is averaged with the thin area, resulting an underestimation of the thickness. This does not happen with samples that are perfectly flat to infinity which are assumed in the simulation. However, we would emphasize that the error of the present method at the thickness range between 200 ~ 800nm is around 10%.

We also measured thickness of the same $SiO_2$ particle by EELS. The mean free path of $SiO_2$ glass is 183 nm[24]. Figures 9(e) and (f) show the thickness distribution and the ratio calculated by dividing the thickness of a sphere with a diameter of 1084 nm by the thickness measured by EELS, respectively. The ratio is around 1.2 and 0.8 at the edge of the particle. To compare the accuracy of 4D-STEM with EELS, the thickness of the sphere whose diameter is 1084 nm was plotted on the vertical axis and the values measured by 4D-STEM (Fig. 9(g)) and EELS (Fig. 9(h)) were plotted on the horizontal axis. In this graph, the point is closer to the diagonal, the more correctly the thickness is measured. When the point is displaced from the diagonal to the upper left, the measurement underestimates the thickness, and when the point is displaced to the lower right, the measurement overestimates the thickness. The thickness of the sphere was plotted on the vertical axis and the values measured by 4D-STEM and EELS were plotted on the horizontal axis. In the case of 4D-STEM, the thickness is underestimated from 200 nm to 800 nm. The amount of underestimate become larger in over 800 nm region, although there are some exceptions in the thin region below about 200 nm. On the other hand, EELS tends to overestimate the thickness in the thin region below about 400 nm, and underestimate it at the position above about 800 nm. The amount of overestimate of EELS is smaller in thick region than the 4D-STEM method. In summary, 4D-STEM method can estimate thickness more accurate than EELS in thin region (less than 800 nm) and less accurate than EELS in thick region (more than 800 nm).

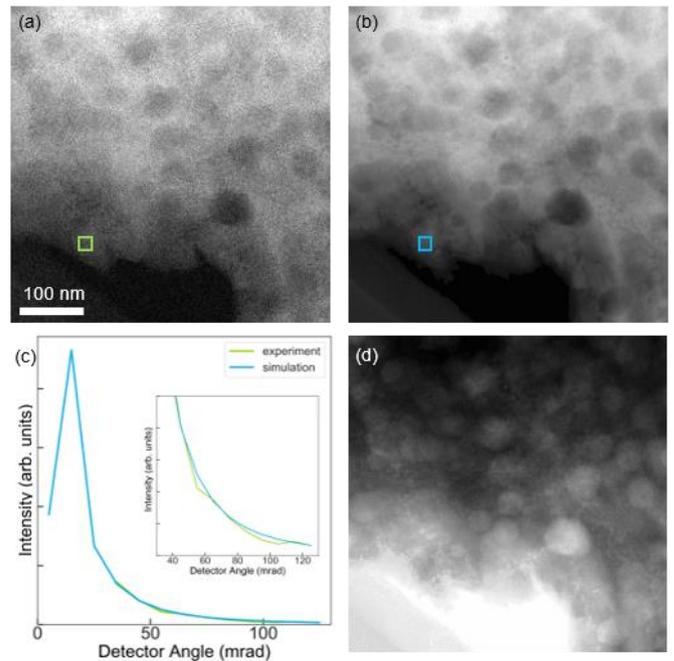

**Figure 10.** (a) Raw HAADF image made by extracting the 120–130 mrad region from the experimentally-obtained AAI. (b) Reproduced HAADF image made by extracting the 120–130 mrad region from AAI designated by the thickness and composition. (c) Comparison of the raw AAI and reproduced AAI. Raw AAI is extracted from the green rectangle in Fig. 9(a) and the reproduced AAI is extracted from the blue rectangle in Fig. 9(b). The inset graph is the enlarged view of the intensity at the high angle region. (d) Reproduced BF image made by extracting the 0–10 mrad region from the AAI designated by the thickness and composition.

Although two constraints, 1) thickness does not abruptly changed and 2) average composition in the observed area is identical to that of the specimen, are required for the present method, this method has a notable benefit for obtaining the composition and thickness maps of electron-beam-sensitive samples. Namely, this method only needs a one-tenth the electron dose of the EELS method and determines the composition with one two-thousandth the electron dose of the EELS method or one-tenth the electron dose of the HAADF and EELS methods.

Further, this method has an additional benefit, an improvement of the image quality. A HAADF image obtained by extracting a 120 to 130 mrad region from the raw 4D-STEM data is shown in Fig. 11(a). The HAADF method only uses highly scattered electrons, which results in a low intensity, statistical fluctuations, and a noisy image. After determining the composition and thickness, the AAI was specified from the AAI group created by simulation using the determined composition and thickness values at each pixel, and the 120 to 130 mrad region was extracted from the AAI to reconstruct a HAADF image. The reconstructed HAADF image is shown in Fig. 10(b). Although the same electron dose was used, the reconstructed HAADF image contained less noise than the HAADF image created from the raw 4D-STEM data. When determining the composition and thickness, we compared the simulated AAI groups with the experimental AAI and chose the composition and thickness which had the most similar AAI as the answer. This operation had the same effect as fitting. Fig. 10 (c) shows a comparison of the AAI obtained from the raw 4D-STEM with the simulated AAI specified by the thickness and composition. The simulated AAI was smoother than the AAI obtained from the raw 4D-STEM. This operation not only denoised but also restored the intensity of the saturated area (0–40 mrad region). Fig. 11(d) shows the bright field (BF) images obtained by extraction of the 0–10 mrad region from the reconstructed AAIs. The BF generated here corresponded to the region where the CCD was saturated and could not be obtained in the experiment. As the contrast of the reconstructed BF was the inverse of the contrast of the HAADF image, the reconstruction of the BF was at least qualitatively correct.

4. Conclusions

We acquired AAIs of the electron diffraction patterns of each scanning point of $27.0BaO-73.0SiO_2$ glass by 4D-STEM. We also simulated diffraction patterns with several thicknesses and compositions of the $BaO-SiO_2$ system. Comparing the simulated AAIs with the rest of the simulated AAIs, we confirmed a one-to-one correspondence between the composition and thickness that reproduced a certain AAI. By comparing the AAIs measured by 4D-STEM with the simulated AAIs, we determined the thickness of the sample with a known composition. Furthermore, we simultaneously determined both the composition and thickness of a sample with $27.0BaO-73.0SiO_2$ under two constraints; 1) thickness does not abruptly change and 2) average composition in the observed area is identical to that of the specimen. Although this method is limited compared with EELS and EDS, in that constraints are required, this method can determine both the thickness and composition with a lower electron dose and a higher accuracy than EELS or EDS. This method has an additional benefit, in an improvement of the quality of the image. It is expected that more various analyses are possible by changing how to apply the constraints.

In this experiment, we have estimated the relative accuracy of the 4D-STEM method by comparing with the EELS and HAADF methods. In future work, we will conduct the 4D-STEM method to investigate the absolute accuracy of this method by measuring a sample with a known thickness and composition.

Acknowledgment

This study was supported by JST-PRESTO (JPM-JPR16NB 16814592) and the Ministry of Education, Culture, Sports, Science and Technology (MEXT); Nos. 17H06094, 19H00818, and 19H05787. STEM-EELS observation was performed at the NIMS microstructural characterization platform as a program of the "Nanotechnology Platform" of MEXT, Japan. The authors acknowledge the helpful support of Dr. Uesugi at NIMS with the experiment. We thank Edanz Group (www.edanzediting.com/ac) for editing a draft of this manuscript.

# Supplementary Information

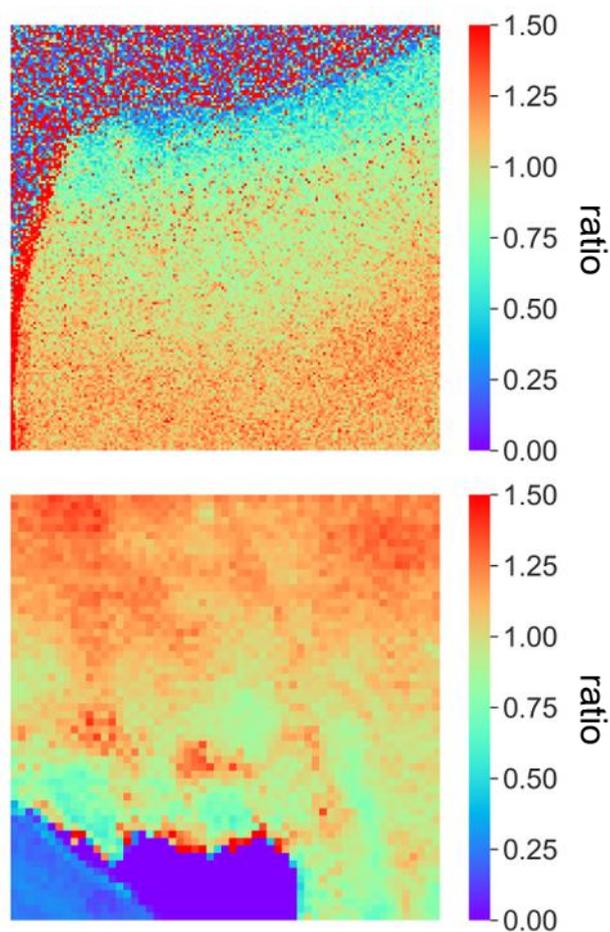

Figure S1. Ratios of thicknesses measured by 4D-STEM and EELS (4D-STEM/EELS). Top image shows the same region as Fig. 5. Bottom image shows the same region as Fig. 6. The 4D-STEM method assesses a thickness that is thinner than the EELS method in the thin area, and thicker than the EELS method in the thick area.

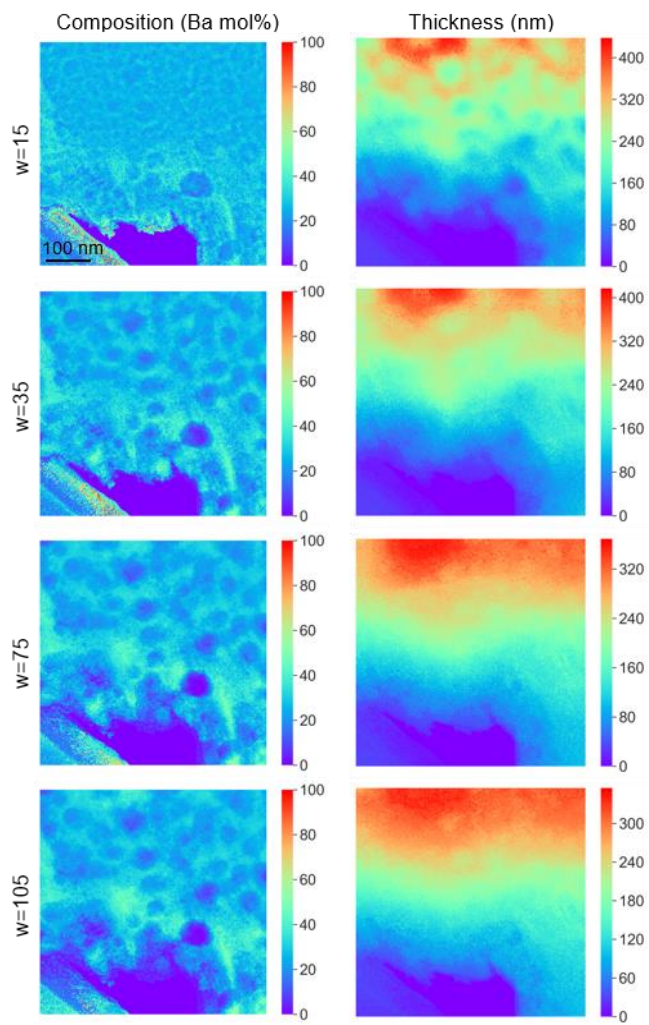

Figure S2. Composition and thickness mapping after updating 14 times with various window sizes. The window sizes are (a) 15, (b) 35, (c) 75 and (d) 105 pixels.

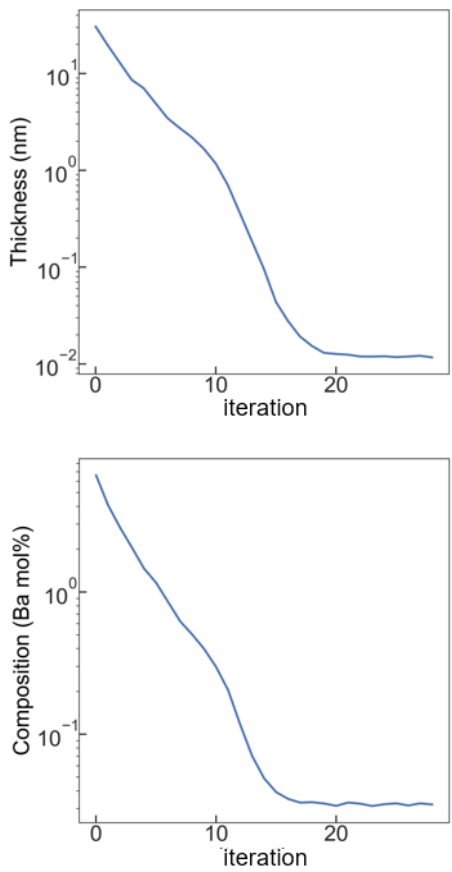

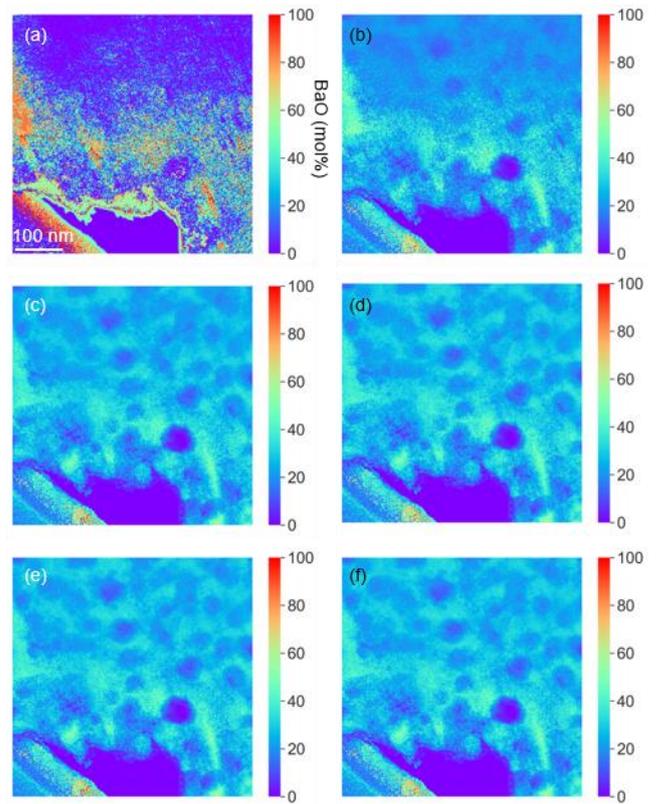

Figure S4. Updating process of the composition. (a) 0, (b) 5, (c) 10, (d) 15, (e) 20 and (f) 25 iterations.

Figure S3. Updating values of the composition and thickness. The updating rate of the composition almost converges under 0.05 after 14 iterations. The updating rate of the thickness also converges under 0.05 after 14 iterations.

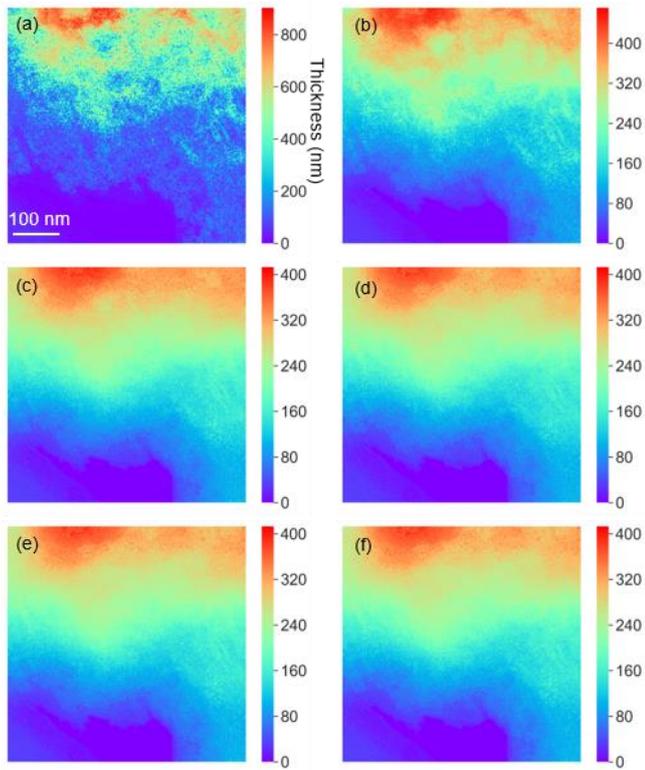

Figure S5. Updating process of the thickness. (a) 0, (b) 5, (c) 10, (d) 15, (e) 20 and (f) 25 iterations.

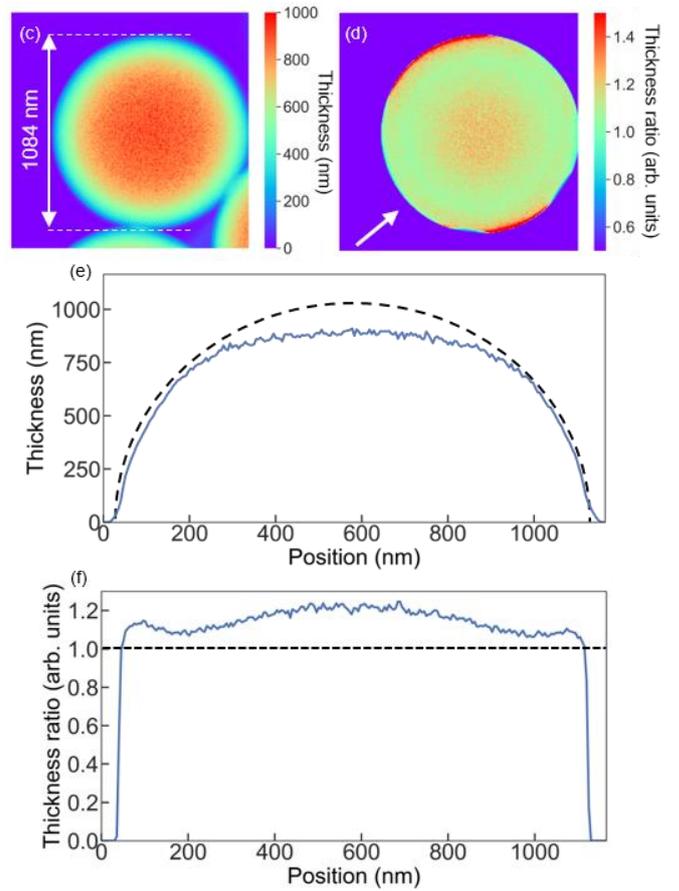

Figure S6. (a) and (b) are the same figures of Fig. 9(c) and (d). Fig. S6(c) and (d) are line profiles of Fig. S6(a) and (b). The line profiles are acquired along the arrow in Fig. S6(b). The black broken line in Fig. S6(c) shows the thickness assuming that the sphere is a perfect sphere with a diameter of 1084 nm. The black broken line in Fig. S6(d) shows the value of thickness ratio is 1.0.